# Strabismic syndromes and syndromic strabismus - a brief review

Aditya Tri Hernowo, B.Sc., M.D., Ph.D[1,2]. and R. Haryo Yudono, M.D., M.Sc.[2]

Department of Basic Medical Sciences, Faculty of Medical and Health Sciences, University Malaysia in Sarawak

Department of Ophthalmology, Faculty of Medicine - Gadjah Mada University

Strabismus can be found in association with congenital heart diseases, for examples, in velocardiofacial (DiGeorge) syndrome, Down syndrome, mild dysmorphic features, in CHARGE association, Turner syndrome, Ullrich-Turner syndrome, cardiofaciocutaneous syndrome.[1-4] Some types of strabismus is heritable (e.g. infantile esotropia syndrome), particularly the ones associated with multisystem disorders, e.g. Moebius syndrome, Prader-Willi syndrome, craniofacial dysostoses, and mitochondrial myopathies.[5] Due to the complexities that a case of strabismus may pertain to, it is worthwhile to get to know more about the strabismic syndromes and syndromic strabismus. This brief review -as its name implied- does not attempt to cover every angle of the syndromic conditions, but offers a refreshment on our knowledge about more prevalent strabismus-related syndromes.

## Duane's syndrome

Duane's syndrome or Duane's retraction syndrome (DRS) is one of the most recognized strabismic syndrome which is genetic in nature. The incidence of DRS is 0.1% in population, and it comprises 1-4% of all strabismus cases.[6] It has a preponderance of left eye involvement[6] and the reason for this is unknown. The gene responsible for DR is located at chromosome 2, 4, 8, and 22[6] (to be more specific: 2q31, 8q13, and 22q11).[7] The occurrence of cleft palate, Klippel-Feil anomaly, sensorineural hearing loss, and DRS are genetically linked.[6] DRS is also associated with velocardiofacial syndrome.[1] From these findings we can conclude that basically the DRS may present in association with certain viscerocranium anomalies.

DRS pathology involves congenital anomaly of the sixth cranial nerve's nuclei (in the form of hypoplasia[8]) with aberrant innervation from the third cranial nerve nuclei (an extra branch of the inferior division of the oculomotor nerve to the lateral rectus[8]).[6] Paradoxical innervation of the lateral rectus muscle causes simultaneous contraction of both medial and lateral recti during adduction, causing the retraction of the globe into the orbita.[6] Both DRS and congenital fibrosis of extraocular muscles can be categorized under incomitant strabismus and fibrosis of the extraocular muscles.[6] In type 1 Duane's syndrome, the abduction saccadic velocity is reduced significantly, while the adduction one is moderately affected.[9] DRS can be associated with hyperopia.[6] Interestingly, despite no significant psychophysical feature changes were identifiable (except for hyperopia), the coronal optic nerve cross-sectional area were reported significantly smaller than normal.[10]

The wide variety of DRS clinicopathologic features demands classification for guiding the management of the condition. The Huber's classification of DRS is: (i) type I - limited abduction with normal to near normal adduction; (ii) type II - limited adduction with normal to near normal abduction; and (iii) type III - limited abduction and adduction.[6] Type I of DRS is further classified based on the deviation in their primary position into: IA - esotropia; IB - exotropia; and IC - orthotropia.[11]

Craniofacial asymmetry can be found in DRS due to long standing torticollis.[6] Thus it is of utmost importance to deal with the syndrome in timely manner. When anomalous head posture was used as a marker for successful surgical intervention, horizontal muscle recession was beneficial in 93% of cases.[12] This procedure is relatively simple, therefore the recognition and timely management of DRS become a determinant in the success of surgery.

## Brown's syndrome

Normally, when the inferior oblique contracts while the globe is adducted, the superior oblique tendon should be relaxed and its tendon passively lengthens. However, this is not the case in Brown's syndrome. Brown's syndrome is also known as superior oblique tendon sheath syndrome, with limited elevation in adduction[13], thus giving an impression (pseudopalsy) of the inferior oblique.[14] CT-scan revealed thickening and inflamed reflected part of the superior oblique.[14]

Brown's syndrome can be hereditary[15] or acquired. The acquired type of the syndrome can be induced by trauma involving orbital throclea.[16] It can also be caused by paranasal sinus mucocele extending into the orbit[17, 18] or orbital venous malformation.[18] Glaucoma implant (Molteno), cataract surgery -due to myotoxicity by local anesthetic agent- (albeit rare), and blepharoplasty may be associated with acquired Brown syndrome.[19, 20, 21] Children with this syndrome often assume an anomalous head posture in the form of head tilt.[22]

Patients with congenital Brown syndrome, may overtime be stabile or resolve spontaneously[23, 24], thus requiring no surgical intervention, especially whose primary position orthotropic (instead of hypotropic).[23] Seventeen percents of the cases present with compensatory head posture, which were completely amenable surgically.[25] Surgical management of this condition usually involves superior oblique tenotomy or tenectomy[26], however, post-operative complication of cyclovertical deviation can be very disturbing for the patients and can be very difficult to treat.[27] In this case, where multiple surgeries failed, partial or sectoral monocular occlusion therapy can alleviate patients with diplopia.[28]

## Möbius' syndrome

The symptoms of Möbius' syndrome can already be identified soon after birth.[29] Several loci were associated with this syndrome: 13q12.2-q13, 3q21-q22, and 10q21.3-q22.[7]

In this syndrome, there is the paralyses of bilateral sixth, seventh[9], and and twelfth cranial nerves.[8] Congenitally, this syndrome may be associated with the loss of innervation to the extraocular muscles, which can also occur in other muscles, hence called congenital innervation dysgenesis syndrome (CID). Histologic examination showed hypoplastic abducens, facial and hypoglossal nerves.[8] Underaction of the lateral rectus in Möbius' syndrome can reach up to -4, whereas the underaction at adduction -3[29], therefore esotropia is one of several features of the Möbius' syndrome.[29] Bilateral talipes equinovarus and syndactyly may be found in association with Möbius' syndrome.[29]

Surgical intervention of choice in both Möbius and Duane's syndrome is recession of the medial rectus with vertical rectus transposition.[29]

## Marcus-Gunn (jaw-winking) phenomenon

Strabismus occur in 36% of Marcus-Gunn syndrome.[30] This syndrome is associated with concomitant esotropia, ptosis, and jaw winking phenomenon.[31] It can also be associated with Williams-Beuren[32] and CFEOM syndrome.[33] Double elevator palsy occured in 25% of the case[34], whereas superior rectus palsy in 23%.[34] The mesencephalic root of the trigeminal nerve (motoric to the mastication muscles), is linked to the muscles of the oculomotor nerve, including to the eyelid levator in this condition.[31] Other aberrant connectivities, such as trigemino-abducens[35] and congenital abnormality within the otolith-oculomotor pathway[36], may accompany the phenomenon.

For the ptosis, bilateral fascial suspension is considered to be the treatment of choice in this syndrome[30, 34, 37], with some advocating unilateral levator excision.[34, 37] Amblyopia occured in 34%[30] to 59%[34] in this syndrome.

## Nystagmus compensation (blockage) syndrome

Around 5% of patients with congenital esotropia had this condition.[38] The syndrome is characterized by early onset esotropia with pseudoparalysis of the abducens nerves. It is called pseudoparalysis because the esotropia is due to convergence (or non-convergence esodeviation[39]), in order to block the nystagmus from occurring (in primary and abduction position).[38] The asymmetric type of this syndrome may manifests itself in the form of asymmetric concomitant horizontal deviations (dissociated horizontal deviation or DHD) dependent on the fixing eye.[40]

Surgical management by recessing the rectus muscle (in this case the medial one) retroequatorially or pre-equatorially[41-44], or with additional resection of the lateral rectus[45], can be employed in the syndrome.[41, 42] While they work for dissociated vertical deviation (DVD), this posterior fixation procedures' results in nystagmus compensation syndrome have been inconclusive, unfortunately.[46]

## Noonan's syndrome

In this congenital heart disease syndrome, strabismus were found in 48% of the cases, of which 80% were horizontal strabismus.[47] Superior oblique tendon absence may also be found[48], this indicate that vertical incomitant strabismus with or without anomalous head posture may be present.

External eye anatomy-related abnormalities in the syndrome include: hypertelorism (74%), downward sloping palpebral apertures (38%), epicanthic folds (39%), ptosis (48%)[47]. Aside from strabismus, the orthoptic findings are: refractive errors (61%), amblyopia (33%), and nystagmus (9%) of cases.[47] Anterior segment involvement include: prominent corneal nerves (46%), anterior stromal dystrophy (4%), cataracts (8%) and panuveitis (2%).[47] Fundal changes occurred in 20% of the study group, including optic nerve head drusen, optic disc hypoplasia, colobomas and myelinated nerves.[47]

## Other syndromes

Other syndromes related to strabismus include congenital fibrosis of the extraocular muscles (CFEOM), fat adherence syndrome, inferior rectus muscle contracture syndrome, and Marfan's syndrome. The syndromes are each briefly described below.

CFEOM showed (from MRI) hypoplasia of the oculomotor, abducens, trochlear nerves, and the extraocular muscles.[8] It can also be associated with agenesis of the corpus callosum, colpocephaly, hypoplasia of the cerebellar vermis, hydrocephalus, pachygyria, encephalocele and/or hydrancephaly.[33] Rarely, it is co-exists with Marcus-Gunn jaw-winking phenomenon.[33]

In fat adherence syndrome restrictive strabismus is present, which is amenable by homologous temporal fascia transplant for globe fixation.[49] The syndrome is an important cause of restrictive strabismus following retina surgery.[50, 51] Another condition would be the inferior rectus muscle contracture syndrome. Its prevalence is 7% according to one study.[52] Under exploration, peribulbar part of the muscle is usually normal, however the retrobulbar part may be fibrotic and showing segmental enlargement in MRI.[53] In one unusual case, spontaneous recovery from the consecutive hypotropia occurred.[54]

Strabismus also occurred in around 11%[55] to 19% of individuals with Marfan syndrome, usually in the form of exotropia, followed by esotropia and other forms.[56] Perhaps due to its nature of connective tissue disorder, media rectus pulley instability was discovered in the syndrome.[57]

## Summary

We have briefly discussed some systemic or regional conditions related to strabismus. Surgery is usually a viable option for treating some conditions above. Nonetheless, when strabismus is a part of a more widespread syndrome (like in Noonan's), or with mesencephalic etiology (like in CFEOM), the treatment becomes difficult, if not impossible at all, and therefore patient education of their condition should be very carefully done.

# Bibliography

1. Mansour AM, Bitar FF, Traboulsi EI, et al. Ocular pathology in congenital heart disease. Eye (Lond) 2005;19(1):29-34.
2. Adhikary HP. Ocular manifestations of Turner's syndrome. Trans Ophthalmol Soc U K 1981;101 (Pt 4):395-6.
3. Weiss E, Loevy H, Saunders A, et al. Monozygotic twins discordant for Ullrich-Turner syndrome. Am J Med Genet 1982;13(4):389-99.
4. Raymond G, Holmes LB. Cardio-facio-cutaneous (CFC) syndrome: neurological features in two children. Dev Med Child Neurol 1993;35(8):727-32.
5. Paul TO, Hardage LK. The heritability of strabismus. Ophthalmic Genet 1994;15(1):1-18.
6. Yuksel D, Orban de Xivry JJ, Lefevre P. Review of the major findings about Duane retraction syndrome (DRS) leading to an updated form of classification. Vision Res 2010;50(23):2334-47.
7. Lorenz B. Genetics of isolated and syndromic strabismus: facts and perspectives. Strabismus 2002;10(2):147-56.
8. Jiao YH, Zhao KX, Wang ZC, et al. Magnetic resonance imaging of the extraocular muscles and corresponding cranial nerves in patients with special forms of strabismus. Chin Med J (Engl) 2009;122(24):2998-3002.
9. Metz HS. Saccadic velocity measurements in strabismus. Trans Am Ophthalmol Soc 1983;81:630-92.
10. Demer JL, Clark RA, Lim KH, Engle EC. Magnetic resonance imaging evidence for widespread orbital dysinnervation in dominant Duane's retraction syndrome linked to the DURS2 locus. Invest Ophthalmol Vis Sci 2007;48(1):194-202.
11. Ahluwalia BK, Gupta NC, Goel SR, Khurana AK. Study of Duane's retraction syndrome. Acta Ophthalmol (Copenh) 1988;66(6):728-30.
12. Barbe ME, Scott WE, Kutschke PJ. A simplified approach to the treatment of Duane's syndrome. Br J Ophthalmol 2004;88(1):131-8.
13. Parks MM, Brown M. Superior oblique tendon sheath syndrome of Brown. Am J Ophthalmol 1975;79(1):82-6.
14. Mafee MF, Folk ER, Langer BG, et al. Computed tomography in the evaluation of Brown syndrome of the superior oblique tendon sheath. Radiology 1985;154(3):691-5.
15. Hamed LM. Bilateral Brown syndrome in three siblings. J Pediatr Ophthalmol Strabismus 1991;28(6):306-9.
16. Legge RH, Hedges TR, 3rd, Anderson M, Reese PD. Hypertropia following trochlear trauma. J Pediatr Ophthalmol Strabismus 1992;29(3):163-6.
17. Pineles SL, Velez FG, Elliot RL, Rosenbaum AL. Superior oblique muscle paresis and restriction secondary to orbital mucocele. J AAPOS 2007;11(1):60-1.
18. Fard MA, Kasaei A, Abdollahbeiki H. Acquired Brown syndrome: report of two cases. J AAPOS 2011;15(4):398-400.
19. Dobler-Dixon AA, Cantor LB, Sondhi N, et al. Prospective evaluation of extraocular motility following double-plate molteno implantation. Arch Ophthalmol 1999;117(9):1155-60.
20. Phillips PH, Guyton DL, Hunter DG. Superior oblique overaction from local anesthesia for cataract surgery. J AAPOS 2001;5(5):329-32.
21. Syniuta LA, Goldberg RA, Thacker NM, Rosenbaum AL. Acquired strabismus following cosmetic blepharoplasty. Plast Reconstr Surg 2003;111(6):2053-9.
22. Boricean ID, Barar A. Understanding ocular torticollis in children. Oftalmologia 2011;55(1):10-26.
23. Kaban TJ, Smith K, Orton RB, et al. Natural history of presumed congenital Brown syndrome. Arch Ophthalmol 1993;111(7):943-6.
24. Tuli S. A 6-year-old girl with restricted upward gaze of her right eye. Pediatr Rev 2012;33(8):e53-6.
25. Kraft SP, O'Donoghue EP, Roarty JD. Improvement of compensatory head postures after strabismus surgery. Ophthalmology 1992;99(8):1301-8.
26. Kerr NC. Management of vertical deviations secondary to other anatomical abnormalities. Am Orthopt J 2011;61:39-48.
27. Santiago AP, Rosenbaum AL. Grave complications after superior oblique tenotomy or tenectomy for Brown syndrome. J AAPOS 1997;1(1):8-15.
28. Routt LA. Monocular partial/sector occlusion therapy: a procedure to inhibit diplopia in Brown syndrome. Optometry 2011;82(4):207-11.
29. Sun LL, Gole GA. Augmented superior rectus muscle transposition for the treatment of strabismus in Mobius syndrome. J AAPOS 2011;15(6):590-2.
30. Doucet TW, Crawford JS. The quantification, natural course, and surgical results in 57 eyes with Marcus Gunn (jaw-winking) syndrome. Am J Ophthalmol 1981;92(5):702-7.
31. Mosavy SH, Horiat M. Marcus Gunn phenomenon associated with synkinetic oculopalpebral movements. Br Med J 1976;2(6037):675-6.
32. Winter M, Pankau R, Amm M, et al. The spectrum of ocular features in the Williams-Beuren syndrome. Clin Genet 1996;49(1):28-31.
33. Pieh C, Goebel HH, Engle EC, Gottlob I. Congenital fibrosis syndrome associated with central nervous system abnormalities. Graefes Arch Clin Exp Ophthalmol 2003;241(7):546-53.
34. Pratt SG, Beyer CK, Johnson CC. The Marcus Gunn phenomenon. A review of 71 cases. Ophthalmology 1984;91(1):27-30.
35. Kodsi S. Marcus Gunn jaw winking with trigemino-abducens synkinesis. J AAPOS 2000;4(5):316-7.
36. Khan AO. Ptotic lid elevation during contralateral head tilt. J AAPOS 2007;11(3):297-9.
37. Bowyer JD, Sullivan TJ. Management of Marcus Gunn jaw winking synkinesis. Ophthal Plast Reconstr Surg 2004;20(2):92-8.
38. von Noorden GK. The nystagmus blockage syndrome. Trans Am Ophthalmol Soc 1976;74:220-36.

39. Dell'Osso LF, Ellenberger C, Jr., Abel LA, Flynn JT. The nystagmus blockage syndrome. Congenital nystagmus, manifest latent nystagmus, or both? Invest Ophthalmol Vis Sci 1983;24(12):1580-7.
40. Zubcov AA, Reinecke RD, Calhoun JH. Asymmetric horizontal tropias, DVD, and manifest latent nystagmus: an explanation of dissociated horizontal deviation. J Pediatr Ophthalmol Strabismus 1990;27(2):59-64; discussion 5.
41. von Noorden GK. Indications of the posterior fixation operation in strabismus. Ophthalmology 1978;85(5):512-20.
42. Reinecke RD. Nystagmus blockage syndrome in the unilaterally blind patient. Doc Ophthalmol 1984;58(1):125-30.
43. Isenberg SJ, Yee RD. The ETHAN syndrome. Ann Ophthalmol 1986;18(12):358-61, 65.
44. Arroyo-Yllanes ME, Fonte-Vazquez A, Perez-Perez JF. Modified Anderson procedure for correcting abnormal mixed head position in nystagmus. Br J Ophthalmol 2002;86(3):267-9.
45. von Noorden GK, Wong SY. Surgical results in nystagmus blockage syndrome. Ophthalmology 1986;93(8):1028-31.
46. Shuckett EP, Hiles DA, Biglan AW, Evans DE. Posterior fixation suture operation (fadenoperation). Ophthalmic Surg 1981;12(8):578-85.
47. Lee NB, Kelly L, Sharland M. Ocular manifestations of Noonan syndrome. Eye (Lond) 1992;6 ( Pt 3):328-34.
48. Sugumaran HK, Pappas JG, Kodsi SR. Congenital absence of the superior oblique tendon in Noonan-neurofibromatosis syndrome. J AAPOS 2011;15(6):593-4.
49. Bagheri A, Erfanian-Salim R, Salour H, Yazdani S. Globe fixation with homologous temporalis fascia transplant for treatment of restrictive esotropia strabismus: an interventional case report and review of the literature. Binocul Vis Strabolog Q Simms Romano 2011;26(4):236-42.
50. Wright KW. The fat adherence syndrome and strabismus after retina surgery. Ophthalmology 1986;93(3):411-5.
51. Hwang JM, Wright KW. Combined study on the causes of strabismus after the retinal surgery. Korean J Ophthalmol 1994;8(2):83-91.
52. Awadein A. Clinical findings, orbital imaging, and intraoperative findings in patients with isolated inferior rectus muscle paresis or underaction. J AAPOS 2012;16(4):345-9.
53. Hamed LM, Mancuso A. Inferior rectus muscle contracture syndrome after retrobulbar anesthesia. Ophthalmology 1991;98(10):1506-12.
54. Sutherland S, Kowal L. Spontaneous recovery from inferior rectus contracture (consecutive hypotropia) following local anesthetic injury. Binocul Vis Strabismus Q 2003;18(2):99-100.
55. Konradsen TR, Zetterstrom C. A descriptive study of ocular characteristics in Marfan syndrome. Acta Ophthalmol 2013.
56. Izquierdo NJ, Traboulsi EI, Enger C, Maumenee IH. Strabismus in the Marfan syndrome. Am J Ophthalmol 1994;117(5):632-5.
57. Oh SY, Clark RA, Velez F, et al. Incomitant strabismus associated with instability of rectus pulleys. Invest Ophthalmol Vis Sci 2002;43(7):2169-78.